\documentclass{siamltex}
\usepackage{epsfig}
\usepackage{amsmath}
\usepackage{amsfonts}
\def\bE{\mathbf E}

\def\bx{\mathbf x}
\def\by{\mathbf y}

\def\bk{\mathbf k}

\def\b0{\mathbf 0}
\def\bTheta{\mathbf \Theta}

\title{Spectrum of the volume integral operator of electromagnetic scattering}

\author{
Neil V. Budko\thanks{Laboratory of Electromagnetic Research, Faculty of Electrical Engineering, Mathematics 
and Computer Science, Delft University of Technology,
Mekelweg~4, 2628~CD, Delft, The Netherlands,
{\tt n.budko@ewi.tudelft.nl}}
\and Alexander B. Samokhin\thanks{Department of Applied Mathematics, Moscow Institute of
Radio Engineering, Electronics, and Automatics (MIREA), Verndasky~av.~78, 117454, Moscow, Russia.}}

\begin{document}

\maketitle

\begin{abstract}
Spectrum of the volume integral operator of the three-dimensional electromagnetic scattering is analyzed. 
The operator has both continuous essential spectrum, which dominates at lower frequencies, and discrete eigenvalues, 
which spread out at higher ones. The explicit expression of the operator's symbol provides exact outline of 
the essential spectrum for any inhomogeneous anisotropic scatterer with H{\"o}lder continuous constitutive parameters. 
Geometrical bounds on the location of discrete eigenvalues are derived for various physical scenarios. 
Numerical experiments demonstrate good agreement between the predicted spectrum of the operator and 
the eigenvalues of its discretized version.
\end{abstract}

\begin{keywords} 
electromagnetics, singular integral operators, symbol, spectrum
\end{keywords}

\pagestyle{myheadings}
\thispagestyle{plain}
\markboth{N.~V.~BUDKO AND A.~B.~SAMOKHIN}{SPECTRUM OF THE VOLUME INTEGRAL OPERATOR}

\section{Introduction}
Knowledge of the spectrum of an integral operator is important for the 
efficient numerical solution of integral equations. For example, it may help to choose an
iterative algorithm, estimate its convergence rate, and construct a proper preconditioner. 
The spectra of electromagnetic integral operators may also find direct application
in physics and engineering: plasma studies, effective 
medium theories, and target identification.

Numerical estimation of the operator's spectrum after discretization is not too 
practical as it requires even more computational resources than the complete 
numerical solution of a scattering problem. Not to mention that such a spectrum 
will be that of a matrix, which only approximates the original operator.
In these circumstances any useful analytical prediction is extremely 
valuable. 

In the pioneering papers \cite{Kleinman1990} and \cite{Kleinman1997} by R.~E.~Kleinman et. al.
the {\it scalar}\, one-, two-, and (acoustical) three-dimensional cases, 
and the {\it surface} (boundary) integral equations were considered. Here we 
look at the integral equation arising in vectorial three-dimensional electromagnetic 
scattering on penetrable objects known as {\it volume} or {\it domain}
integral equation.

First results concerning the spectrum of the three-dimensional
electromagnetic scattering problems were derived from the analytical 
Mie solution for a spherical scatterer. This case was analyzed in considerable 
detail in \cite{Gastine1967}, \cite{Conwell1984}, \cite{Hunter1988}, and the
so-called Mie-resonances were identified as eigenvalues.
In \cite{Rahola2000} J.~Rahola has compared the theory 
of the Mie resonances with the numerical spectrum of the volume integral operator. 
Numerical experiments indicated, though, that apart from the expected discrete Mie resonances, the spectrum of the discretized 
volume integral operator contained eigenvalues clustered along distinct line segments in 
the complex plane. However, the {\it ad hoc} explanation of this 
phenomenon presented in \cite{Rahola2000} cannot be extended on the case of a general 
inhomogeneous scatterer. The latter requires a different and more rigorous approach. 

Our goal is to establish the
connection between the operator's spectrum and the physical properties of a scatterer. Using the symbol of the 
electromagnetic volume integral operator we show that in addition to discrete eigenvalues 
the spectrum has an essential continuous part as well. Fortunately, the symbol can be derived explicitly allowing
for exact predictions of the essential spectrum for any H{\"o}lder continuous anisotropic scatterer.
In the isotropic case the essential spectrum is just the set of values admitted by the constitutive
parameters of the object in ${\mathbb R}^{3}$. 
The approximate location of the discrete eigenvalues is estimated in 
the spirit of R.~E.~Kleinman et. al., and depends mostly on the extent and shape of the object.

The paper is organizer as follows. In the next two sections we recall several more or less well-known facts
from the theory of Fredholm operators with pointwise multipliers and simple singular integral operators. Driven by the electromagnetic 
problem at hand in Section~4 we
consider `combined' operators consisting of both pointwise multiplication 
operators and simple singular integral operators. Theory of such combined operators was 
developed by Mikhlin et. al. \cite{Mikhlin1}, \cite{Mikhlin2} and we hope that 
our brief introduction will help to popularise their approach. The main purpose of these three sections is to
introduce the concept of a symbol of a combined operator, which helps to establish the
essential part of the operator's spectrum. Having mustered this tool we move on to the electromagnetic case where
we construct the symbol of the volume integral operator, analyze the essential spectrum, and provide
geometrical bounds on the distribution of discrete eigenvalues. Finally, we present some numerical experiments
with the matrix version of the operator, and try to comprehend similarities and differences
between the continuous and the discrete worlds.

\section{Spectrum of Fredholm operators with pointwise multipliers}
The volume scattering in one and two dimensions is described by integral equations
with manifestly Fredholm operators, i.e. an identity operator plus a compact integral 
operator. This is also the case in the acoustic three-dimensional scattering
and has to do with the fact that the singularity in the kernel of all these 
integral operators is weak, i.e. its order is smaller than three (dimension of the spatial manifold),
whereas, the support of the scatterer is finite.
For integral equations with Fredholm operators the questions of existence and uniqueness
are linked, so that it is sufficient to establish the uniqueness of the solution only.

With operators which are not of the form `identity plus compact' the Fredholm property
is established by proving the existence of a regularizer - an operator which, when applied to 
the operator in question, reduces it to the form `identity plus compact'. There may be a left 
and a right regularizer, naturally. First, we clarify one important thing: the spectra of the
original Fredholm operator and its regularized version are, in general, different.
Indeed, consider the following Fredholm integral 
equation:
 \begin{align}
 \label{eq:AlmostFred}
 a(\bx) u(\bx) + \int\limits_{\bx, \bx'\in D}K(\bx,\bx')u(\bx')\;{\rm d}\bx' = v(\bx)\;,
 \end{align}
where instead of the identity operator we have a (pointwise) multiplication with 
some continuous function $a(\bx)\ne 0$, $\bx\in D$. The integral operator is presumed to be compact.
In operator notation we shall write
 \begin{align}
 \label{eq:AlmostFredOp}
 \left[{\rm d}(a) + K\right] u = v\;,
 \end{align}
where ${\rm d}(a)$ indicates that we have a multiplication operator, analogous
to a diagonal matrix.
The reqularized Fredholm equation is obtained from (\ref{eq:AlmostFred}) by a trivial 
operation of (pointwise) multiplication by $a(\bx)^{-1}$, i.e.
 \begin{align}
 \label{eq:TrueFred}
 u(\bx) + a(\bx)^{-1}\int\limits_{\bx,\bx'\in D}K(\bx,\bx')u(\bx')\;{\rm d}\bx' = 
 a(\bx)^{-1}v(\bx).
 \end{align}
Hence, the regularizer in this case is simply ${\rm d}(a)^{-1}$, as it gives the
`identity plus compact' form
 \begin{align}
 \label{eq:TrueFredOp}
 {\rm d}(a)^{-1}\left[{\rm d}(a) + K\right]=I + {\rm d}(a)^{-1}K=I + K',
 \end{align}
where $K'$ is a compact operator. The spectra of the two 
(original and regularized) operators are quite different though.
Indeed, consider operator
 \begin{align}
 \label{eq:ResolvOpScalar}
 {\rm d}(a)-\lambda I + K\;,
 \end{align}
and let $\lambda=a(\bx_{0})$, for some arbitrary $\bx_{0}\in D$.
If $\lambda$ does not belong to the spectrum, then there exists a bounded inverse 
of (\ref{eq:ResolvOpScalar}) for that particular $\lambda$. This means that
if we consider equality
 \begin{align}
 \label{eq:ResolvEqScalar}
 \left[{\rm d}(a)-\lambda I + K\right]u=v\;,
 \end{align}
then for all possible perturbations $\Delta v$ of the right-hand side  and the corresponding perturbations
$\Delta u$ of the solution there exists a constant $C<\infty$ such that
 \begin{align}
 \label{eq:Bounded}
 \Vert \Delta u\Vert\le C\Vert\Delta v\Vert.
 \end{align}
Due to the contunuity of $a(\bx)$ in $D$ we can always choose a neighborhood 
of $\bx_{0}$, say $\delta(\bx_{0})$, such that $\vert a(\bx)-a(\bx_{0})\vert<\epsilon$, for $\bx\in\delta(\bx_{0})$.
Consider perturbations $\Delta u(\bx)$ which are zero outside $\delta(\bx_{0})$ and have a fixed norm $\Vert \Delta u\Vert=R$,
independent of $\delta(\bx_{0})$. This can be a function of the following type:
 \begin{align}
 \label{eq:Perturb}
 \Delta u (\bx) \sim e^{i\bx\cdot\bk(\delta,\epsilon)},\;\;\;\; \bx\in \delta(\bx_{0}).
 \end{align}
or a similar oscillating function with varying spatial frequency $\bk(\delta,\epsilon)$. Moreover, since $K$ is compact,
we may assume that our perturbations give $\Vert K\Delta u\Vert\le \kappa\epsilon$, for all sufficiently small $\delta(\bx_{0})$.
Further, we have $\Vert[a(\bx)-a(\bx_{0})]\Delta u\Vert\le R\epsilon$. Hence, for the perturbation of the right-hand side we obtain
 \begin{align}
 \label{eq:TheBound}
 \left\Vert\Delta v\right\Vert \le \left\Vert(a-\lambda)\Delta u \right\Vert
 +\left\Vert K \Delta u\right\Vert \le (R + \kappa) \epsilon\,,
 \end{align}
whereas, $\Vert\Delta u\Vert=R$ for all $\Delta u$. For the condition (\ref{eq:Bounded}) to hold we must 
be able to provide $C<\infty$ such that
 \begin{align}
 \label{eq:CondBounded}
 R\le C (R+\kappa)\epsilon\,,
 \end{align}
for all $\epsilon$, and this is obviously not possible, since we can make $\epsilon$ as small as we 
like by reducing $\delta(\bx_{0})$, while keeping $R$ constant. Hence, the inverse of operator (\ref{eq:ResolvOpScalar}) is
unbounded and $\lambda=a(\bx_{0})$ belongs to the spectrum of operator $[{\rm d}(a)+K]$.

If continuous function $a(\bx)$ is not constant on $D$, then the set of its values is an extended dense subset 
of the complex plane. Such a spectrum is called continuous as opposed to the usual discrete eigenvalues.
In addition to continuous spectrum operator $[{\rm d}(a)+K]$ may have discrete eigenvalues as well.
The regularized operator $[I+K']$, however, is known to have discrete spectrum only. 
This follows from the fact that the spectrum of $K'$ is purely discrete \cite{Kreyszig}. Thus, 
regularization destroys continuous spectrum. Finally, we notice that operator (\ref{eq:ResolvOpScalar})
cannot be regularized at the points of continuous spectrum. 

\section{Spectrum of simple singular integral operators}
The operator of the three-dimensional electromagnetic scattering has the following form
 \begin{align}
 \label{eq:StrongGeneral}
 Au={\rm d}(a)u+ A^{\rm s}{\rm d}(b)u + K u\,,
 \end{align}
 where $K$ is compact and
 \begin{align}
 \label{eq:ScalarSingOp}
 A^{\rm s}{\rm d}(b)u=\lim\limits_{\epsilon\rightarrow 0} \int\limits_{\bx'\in{\mathbb R}^{3}\setminus\delta(\epsilon)}
 \frac{f(\bTheta)}{r^{3}}b(\bx')u(\bx')\,{\rm d}\bx'\,,
 \end{align}
with $r=\vert\bx'-\bx\vert$, and $\bTheta=(\bx'-\bx)/r$. The principal-value integral above
is understood as an integral over entire ${\mathbb R}^{3}$ excluding a ball around $\bx$ whose
radius tends to zero in the limit. This integral
exists under certain assumptions on $b(\bx)u(\bx)$ and $f(\bTheta)$. The necessary condition is the following one:
 \begin{align}
 \label{eq:FCond}
 \int\limits_{\bTheta\in {\mathbb S}} f(\bTheta){\rm d}S =0\,,
 \end{align}
where integration is performed over a unit sphere ${\mathbb S}$. The H{\"o}lder-continuity of $b(\bx)u(\bx)$ in ${\mathbb R}^{3}$
is another (sufficient) condition. Operator $A^{\rm s}$  in (\ref{eq:ScalarSingOp}) is called the {\it simple singular integral operator}.
Its kernel is not an integrable function, and the best way to comprehend the meaning of such a $p.v.$-integral is to take the 
three-dimensional Fourier transform of $A^{\rm s}u$. Like this
 \begin{align}
 \label{eq:FTDerive}
 \begin{split}
 \tilde{v}(\bk)&=\lim\limits_{\epsilon\rightarrow 0}
 \int\limits_{\bx\in{\mathbb R}^{3}}
 \int\limits_{\bx'\in{\mathbb R}^{3}\setminus\delta(\epsilon)} \frac{f(\bTheta)}{r^{3}}u(\bx')
 e^{-i\bk\cdot\bx}{\rm d}\bx'{\rm d}\bx
 \\
 &=\lim\limits_{\epsilon\rightarrow 0}
 \int\limits_{\bx\in{\mathbb R}^{3}}
 \int\limits_{\bx'\in{\mathbb R}^{3}\setminus\delta(\epsilon)} \frac{f(\bTheta)}{r^{3}}u(\bx')
 e^{-i\bk\cdot(\bx-\bx')}e^{-i\bk\cdot\bx'}{\rm d}\bx'{\rm d}\bx
 \\
 &=
 \lim\limits_{\epsilon\rightarrow 0}
 \int\limits_{\by\in{\mathbb R}^{3}\setminus\delta(\epsilon)} \frac{f(-\bTheta)}{r^{3}}
 e^{-i\bk\cdot\by}{\rm d}\by
 \int\limits_{\bx'\in{\mathbb R}^{3}} u(\bx')e^{-i\bk\cdot\bx'}{\rm d}\bx'
 \\
 &=
  F(\bk)\tilde{u}(\bk)\,.
 \end{split}
 \end{align}
In other words, we have arrived at a familiar $\bk$-domain expression for the Fourier transform of a convolution.
The function $F(\bk)$ can only be found, if one knows the explicit form of $f(\bTheta)$. However,
there is something to say even in the general case. Consider this
 \begin{align}
 \label{eq:DeriveFk}
 \begin{split}
 F(\bk)&= \lim\limits_{\epsilon\rightarrow 0}
 \int\limits_{\by\in{\mathbb R}^{3}\setminus\delta(\epsilon)} \frac{f(-\bTheta)}{r^{3}}
 e^{-i\bk\cdot\by}{\rm d}\by
 \\
 &=\lim\limits_{\epsilon\rightarrow 0}\int\limits_{\epsilon}^{\infty}\int\limits_{S}
 \frac{f(-\bTheta)}{r}e^{-ir\rho\cos\gamma}{\rm d}S{\rm d}r
 =\lim\limits_{\epsilon\rightarrow 0}\int\limits_{\rho\epsilon}^{\infty}\int\limits_{S}
 f(-\bTheta)\frac{e^{-it\cos\gamma}}{t}{\rm d}S{\rm d}t
 \\
 &=\lim\limits_{\epsilon\rightarrow 0}\int\limits_{\rho\epsilon}^{1}\int\limits_{S}
 f(-\bTheta)\frac{e^{-it\cos\gamma}}{t}{\rm d}S{\rm d}t
 +\int\limits_{1}^{\infty}\int\limits_{S}
 f(-\bTheta)\frac{e^{-it\cos\gamma}}{t}{\rm d}S{\rm d}t
 \\
 &=\int\limits_{1}^{\infty}\int\limits_{S}
 f(-\bTheta)\frac{e^{-it\cos\gamma}}{t}{\rm d}S{\rm d}t
 +\lim\limits_{\epsilon\rightarrow 0}\int\limits_{\rho\epsilon}^{1}\int\limits_{S}
 f(-\bTheta)\frac{1}{t}\left[e^{-it\cos\gamma}-1\right]{\rm d}S{\rm d}t
 \\
 &\;\;\;\;\;\;+\lim\limits_{\epsilon\rightarrow 0}\int\limits_{\rho\epsilon}^{1}\frac{1}{t}{\rm d}t
 \int\limits_{S} f(-\bTheta){\rm d}S
 \\
 &=\int\limits_{1}^{\infty}\int\limits_{S}
 f(-\bTheta)\frac{e^{-it\cos\gamma}}{t}{\rm d}S{\rm d}t
 +\int\limits_{0}^{1}\int\limits_{S}
 f(-\bTheta)\frac{1}{t}\left[e^{-it\cos\gamma}-1\right]{\rm d}S{\rm d}t
 \end{split}
 \end{align}
Above we have itroduced spherical coordinate systems for 
both $\by$ and $\bk$, where $\rho=\vert\bk\vert$, and used condition (\ref{eq:FCond}). 
As one can see, since $\epsilon\rightarrow 0$ tends to zero independently of $\rho$,
$F(\bk)$ does not depend on $\rho$ and, hence, it is independent of $\vert\bk\vert$. 
In other words, if $F(\bk)$ depends on $\bk$ at all, then it is only the direction of $\bk$,
not its magnitude, i.e.
 \begin{align}
 \label{eq:FkDependence}
 F(\bk)=F\left(\frac{\bk}{\vert\bk\vert}\right)=F(\tilde{\bTheta})\,.
 \end{align}
Note that this is only true if the order of singularity in (\ref{eq:ScalarSingOp}) is exactly three.
If we had a weakly singular integral, then $F(\bk)$ would depend on $\vert\bk\vert$
explicitly. 

Let us now construct a regularizer for a simple singular operator $A^{\rm s}$. 
Denoting by ${\mathcal F}$ and ${\mathcal F}^{-1}$, correspondingly the forward and the inverse
Fourier transforms,  and by $F(\tilde{\bTheta})^{-1}$ the inverse of  $F(\tilde{\bTheta})$, 
we can do, for example, the following:
 \begin{align}
 \label{eq:RegSimple}
 {\mathcal F}^{-1}F(\tilde{\bTheta})^{-1}{\mathcal F}A^{\rm s}u=
 {\mathcal F}^{-1}F(\tilde{\bTheta})^{-1}F(\tilde{\bTheta}){\mathcal F}u=
 {\mathcal F}^{-1}{\mathcal F}u=u\,.
 \end{align}
In this way the original operator $A^{\rm s}$ is reduced by the regularizer 
${\mathcal F}^{-1}F(\tilde{\bTheta})^{-1}{\mathcal F}$ to an identity operator, i.e.
the regularizer is also an inverse of the operator. 

To recover the spectrum of a simple singular operator we need to consider 
 \begin{align}
 \label{eq:ResolventSingOp}
 \left[A^{\rm s}-\lambda I\right] u=\lim\limits_{\epsilon\rightarrow 0} \int\limits_{\bx'\in{\mathbb R}^{3}\setminus\delta(\epsilon)}
 \frac{f(\bTheta)}{r^{3}}u(\bx')\,{\rm d}\bx'
 -\lambda u(\bx)\,.
 \end{align}
Repeating calculations (\ref{eq:FTDerive}) we arrive at
the following equivalent version of the spectral problem:
 \begin{align}
 \label{eq:ResolvTransf}
 F(\tilde{\bTheta}) \tilde{u}(\bk) = \lambda\tilde{u}(\bk)\,,
 \end{align}
Hence, generalizing this to matrix-valued $F(\tilde{\bTheta})$, we conclude that the
domain $s(\lambda)$ in the complex plane consisting of all eigenvalues $\lambda$
of matrix $F(\tilde{\bTheta})$ for all $\tilde{\bTheta}$ 
on the unit sphere belongs to the spectrum of the simple singular integral 
operator $A^{\rm s}$.
Note, although matrix eigenvalues are always discrete, they may vary continuously with $\tilde{\bTheta}$, 
thus creating a continuous spectrum. Again, the operator of (\ref{eq:ResolventSingOp}) cannot be regularized
for $\lambda\in s(\lambda)$.

\section{Essential spectrum of combined operators}
In the previous two sections we saw that the spectrum and the regularization of operators 
are tightly linked. Namely, operators of the spectral problems (\ref{eq:ResolvOpScalar}) and (\ref{eq:ResolventSingOp}) could not be 
regularized for $\lambda\in s(\lambda)$. This ia also true, if we consider the spectral problem for operator 
(\ref{eq:StrongGeneral}), and, perhaps, for any operator. However, a rigorous proof of this link for a
general case is too abstract, and we shall not provide it here. Taking it for granted we shall instead concentrate 
on the construction of a general regularizer.

We have shown that the Fredholm operator with 
free term ${\rm d}(a)$ could be regularized by ${\rm d}(a)^{-1}$, whereas 
a simple singular integral operator is regularized by 
${\mathcal F}^{-1}F(\tilde{\bTheta})^{-1}{\mathcal F}$.
However, neither of these approaches works with the general 
operator (\ref{eq:StrongGeneral}) arising in the electromagnetic scattering problem. 
Since this general operator is a combination
of multiplication and simple singular operators (further -- {\it combined} operator) 
we expect that a proper combination
of the two approaches may work.

Technique presented below is a simplified version of the approach suggested 
by S.~G.~Mikhlin \cite{Mikhlin1}, \cite{Mikhlin2}. 
First thing to note is that, while searching for a regularizer, the actual form of the compact 
operator in the resulting `identity plus compact' expression is not important.
Secondly, a regularizer cannot be a compact operator, as the product of a compact operator 
with any other bounded linear operator is a compact operator. Finally, the product of any
two combined operators is also a combined operator (plus a compact one, in general). 
For example the product of two operators of type (\ref{eq:StrongGeneral})
will look like
 \begin{align}
 \label{eq:ProductGeneral}
 \begin{split}
 A_{1}A_{2}&=\left[{\rm d}(a_{1})+ A^{\rm s}_{1}{\rm d}(b_{1}) + K_{1} \right]
 \left[{\rm d}(a_{2})+ A^{\rm s}_{2}{\rm d}(b_{2}) + K_{2} \right]
 \\
 &={\rm d}(a_{1}){\rm d}(a_{2})+{\rm d}(a_{1})A^{\rm s}_{2}{\rm d}(b_{2})+
 A^{\rm s}_{1}{\rm d}(b_{1}){\rm d}(a_{2})+A^{\rm s}_{1}{\rm d}(b_{1})A^{\rm s}_{2}{\rm d}(b_{2}) + K\,,
 \end{split}
 \end{align}
where all compact operators are hidden here in the generic compact operator $K$.

To every combined operator there corresponds a function called {\it symbol}
constructed in accordance with the following set of rules.

\begin{definition}
\label{def:Symbol}
\begin{enumerate} 
\item{The symbol of a compact operator is zero.}
\item{The symbol of a pointwise multiplication operator with a H{\"o}lder continuous multiplier 
is the multiplier itself.}
\item{The symbol of a simple singular integral operator is its Fourier transform.}
\item{The symbol of the sum (product) of the aforementioned operators is the sum (product) 
of their symbols (the order of symbols in a product corresponding to the order
of operators).}
\end{enumerate}
\end{definition}
For the last property to hold it is sufficient to ensure that the multipliers are
H{\"o}lder-continuous functions in ${\mathbb R}^{3}$, so that the products of simple singular integral 
operators and multiplication operators and the products of their symbols are well defined. 
According to Definition~\ref{def:Symbol}, the symbol of the combined 
operator (\ref{eq:StrongGeneral}) is just this function
 \begin{align}
 \label{eq:GenSymb}
 {\rm Symb}\left[A\right]=\Phi(\bx,\tilde{\bTheta})=a(\bx)+F(\tilde{\bTheta})b(\bx)\,,
 \end{align}
and the symbol of the product (\ref{eq:ProductGeneral}) of 
two combined operators is
 \begin{align}
 \label{eq:GenProdSymb}
 \begin{split}
 {\rm Symb}\left[A_{1}A_{2}\right]&
 =a_{1}(\bx)a_{2}(\bx)+a_{1}(\bx)F_{2}(\tilde{\bTheta})b_{2}(\bx)+
 F_{1}(\tilde{\bTheta})b_{1}(\bx)a_{2}(\bx)
 \\
 &+F_{1}(\tilde{\bTheta})b_{1}(\bx)
 F_{2}(\tilde{\bTheta})b_{2}(\bx) =\Phi_{1}(\bx,\tilde{\bTheta})\Phi_{2}(\bx,\tilde{\bTheta})
 \\
 &={\rm Symb}\left[A_{1}\right]{\rm Symb}\left[A_{2}\right]\,.
 \end{split}
 \end{align}
For our purposes the most important property of a symbol is that
the correspondence between a combined operator and its symbol is one-to-one
up to a compact operator. This means that to each combined operator there
corresponds only one symbol, whereas to each symbol there may correspond 
more than one operator, but their difference would be a compact operator.

Suppose now that we want to find a regularizer $A^{\rm r}$ of a combined operator  $A$, i.e.
 \begin{align}
 \label{eq:RegStrongGeneral}
 A^{\rm r}A=A^{\rm r}\left[{\rm d}(a)+A^{\rm s}{\rm d}(b) +K\right]=I+K'\,.
 \end{align}
The symbol of this product can be constructed using Definition~\ref{def:Symbol} and
is given by
 \begin{align}
 \label{eq:SymbRegStrongGeneral}
 {\rm Symb}\left[A^{\rm r}A\right]={\rm Symb}\left[A^{\rm r}\right]\left[a(\bx)+F(\tilde{\bTheta})b(\bx)\right]=1\,.
 \end{align}
To this symbol there corresponds an
operator of the form `identity plus compact', i.e. should we succeed in obtaining such a unit symbol --
the regularization is accomplished.
Therefore, the symbol of a regularizer, say $\Phi_{\rm r}(\bx,\tilde{\bTheta})$, must be an inverse of the 
symbol of the combined operator in question. Since the latter symbol may be a matrix-valued function, one may have to 
find an inverse of a matrix.

In this simplified version of the Mikhlin's approach summarized in Definition~\ref{def:Symbol} 
we have limited ourselves to combined operators. Therefore, in order to be able to apply the 
multiplication property in (\ref{eq:SymbRegStrongGeneral}) we have to be sure
that the regularizer, if it exists, belongs to the class of combined operators as well. 
This would be the case if the inverse of the symbol of a combined operator -- $[a(\bx)+F(\tilde{\bTheta})b(\bx)]^{-1}$ --
had the form of a symbol of a combined operator itself, i.e. contained only
multipliers and Fourier transforms of simple singular integral operators. 
Fortunately, it follows from the Cayley-Hamilton theorem for matrices that the 
inverse of a nonsingular matrix can always be represented as a polynomial of this  
matrix. Obviously, any such polynomial contains only the products of the multipliers and
the Fourier transforms mentioned above.

Conditions on the existence of a regularizer thus become conditions
on the existence and boundedness of the inverse of the symbol of a combined operator 
for all $\bx\in{\mathbb R}^{3}$ and all $\tilde{\bTheta}$
on a unit sphere.

The symbol will help us to recover the essential spectrum of combined operator $A$ given by (\ref{eq:StrongGeneral}).
Since operator $A-\lambda I$ is also a combined operator, its symbol is given by
 \begin{align}
 \label{eq:SymbResolv}
 \Phi(\bx,\tilde{\bTheta},\lambda)=a(\bx)+F(\tilde{\bTheta})b(\bx)-\lambda{\mathbb I}=\Phi(\bx,\tilde{\bTheta})-\lambda{\mathbb I}\,,
 \end{align}
where ${\mathbb I}$ is an identity matrix of the corresponding dimensions.
This symbol is a parametric function of $\lambda$ as well as $\bx$ and $\tilde{\bTheta}$. 
Suppose that $\lambda$ coinsides with the 
eigenvalue of the symbol matrix $\Phi(\bx,\tilde{\bTheta})$ for some $\bx$ and $\tilde{\bTheta}$. 
Then, operator $A-\lambda I$ cannot be regularized and, hence, does not have a bounded inverse.
In summary (generalizing to matrix-valued symbols):
domain $s(\lambda)$ in the complex plane consisting of all eigenvalues 
of symbol $\Phi(\bx,\tilde{\bTheta})$ for 
all $\bx\in{\mathbb R}^{3}$ and all $\tilde{\bTheta}\in{\mathbb S}$ 
belongs to the (essential) spectrum of the corresponding combined operator $A$. 
If the eigenvalues of $\Phi(\bx,\tilde{\bTheta})$
vary continuously in the complex plane with $\bx$ and $\tilde{\bTheta}$ varying correspondingly
in ${\mathbb R}^{3}$ and on ${\mathbb S}$, then the essential spectrum is continuous.

\section{The electromagnetic case}
Now we turn to the electromagnetic three-dimensional case.
In the frequency domain with time-factor 
$\exp(-i\omega t)$ the total electric field $\bE$ in the presence of an
inhomogeneous anisotropic electric-type scatterer satisfies the following singular
integral equation (see e.g. \cite{SamokhinBook})
 \begin{align}
 \label{eq:EFVIE}
 \begin{split}
 \bE^{\rm in}(\bx,\omega)
 = 
 &
 \left[{\mathbb I} + \frac{1}{3}\chi(\bx,\omega)\right]
 \bE(\bx,\omega)
 \\
 & - \lim\limits_{\epsilon\rightarrow 0}
 \int\limits_{\bx'\in {\mathbb R}^{3}\setminus\vert\bx-\bx'\vert<\epsilon}
 {\mathbb G}_{0}(\bx-\bx')
 \chi(\bx',\omega)
 \bE(\bx',\omega)
 \;{\rm d}\bx'
 \\
 &- \int\limits_{\bx'\in {\mathbb R}^{3}}
 {\mathbb G}_{1}(\bx-\bx',\omega)
 \chi(\bx',\omega)
 \bE(\bx',\omega)
 \;{\rm d} \bx'
 \end{split}
 \end{align}
where $\bE^{\rm in}$ is a given incident electric field, and $\chi$
is the tensor of relative electric contrast of the object with respect to a 
homogeneous isotropic background. Explicitly it is given by
 \begin{align}
 \label{eq:ChiTen}
 \chi(\bx,\omega)=\frac{1}{\eta_{\rm b}(\omega)}\eta(\bx,\omega)-{\mathbb I}\;,
 \end{align}
where ${\mathbb I}$ denotes the $3\times 3$ identity matrix and the constitutive parameters of 
the scatterer are described by the tensor of the transverse medium admittance (per length),
which in its turn contains the more familiar tensors of conductivity $\sigma(\bx,\omega)$ and dielectric 
permittivity $\varepsilon(\bx,\omega)$, namely,
 \begin{align}
 \label{eq:EtaTen}
 \eta(\bx,\omega)=\sigma(\bx,\omega)-i\omega\varepsilon(\bx,\omega)\;.
 \end{align}
The {\it isotropic} background medium has parameters 
$\eta_{\rm b}(\omega)=\sigma_{\rm b}(\omega)-i\omega\varepsilon_{\rm b}(\omega)$ 
and $\zeta_{\rm b}(\omega)=-i\omega\mu_{\rm b}(\omega)$, where $\mu_{\rm b}$
is the magnetic permeability. The so-called wavenumber $k_{\rm b}(\omega)$ of this medium 
can be found from $k^{2}_{\rm b}(\omega)=-\eta(\omega)\zeta(\omega)$ by taking a proper branch of
a square root. Although the contrast function is different from zero only within a finite
spatial domain $D$, we have extended the limits of integration in (\ref{eq:EFVIE}) to the whole
of ${\mathbb R}^{3}$. This formal extention allows direct application of the results of
the previous section. 

The kernels of the intergal operators in (\ref{eq:EFVIE}) contain tensors ${\mathbb G}_{0}$ 
and ${\mathbb G}_{1}$, the first of which is given by
 \begin{align}
 \label{eq:Gten0}
 \begin{split}
 {\mathbb G}_{0}(\bx) = 
 \frac{1}{4\pi\vert\bx\vert^{3}}
 \left[3\,{\mathbb Q} - {\mathbb I}\right],
 \end{split}
 \end{align}
where tensor ${\mathbb Q}=\bTheta(\bTheta\cdot\;)$ can also be written in matrix notation
as ${\mathbb Q}=\bTheta\bTheta^{T}$. This tensor is
an orthogonal projector, i.e. ${\mathbb Q}^{2}=\bTheta\bTheta^{T}\bTheta\bTheta^{T}={\mathbb Q}$.
The second kernel contains tensor
 \begin{align}
 \label{eq:Gten1}
 \begin{split}
 {\mathbb G}_{1}(\bx,\omega)= 
 \left[
 - \alpha(\bx,\omega)
 - 3\beta(\bx,\omega)
 + 3\gamma(\bx,\omega)
 \right]{\mathbb Q}
 +
 \left[
   \alpha(\bx,\omega)
 + \beta(\bx,\omega)
 - \gamma(\bx,\omega)
 \right]{\mathbb I},
 \end{split}
 \end{align}
with 
 \begin{align}
 \label{eq:AlpBetGam}
 \begin{split}
 \alpha(\bx,\omega) = k_{\rm b}^{2}(\omega) 
 \frac{e^{i k_{\rm b}(\omega)\vert\bx\vert}}
 {4\pi\vert\bx\vert},
 \;\;\;\;
 \beta(\bx,\omega) = i\,k_{\rm b}(\omega)
 \frac{e^{i k_{\rm b}(\omega)\vert\bx\vert}}
 {4\pi\vert\bx\vert^{2}},
 \;\;\;\;
 \gamma(\bx,\omega) =
 \frac{e^{i k_{\rm b}(\omega)\vert\bx\vert}-1}
 {4\pi\vert\bx\vert^{3}}.
 \end{split}
 \end{align}

In operator notation of the previous section our equation is written as
 \begin{align}
 \label{eq:EFVIEOp}
 \left[{\rm d}({\mathbb I}+\frac{1}{3}\chi) + A^{\rm s}{\rm d}({\chi}) + K\right] u = u^{\rm in}\;.
 \end{align}
The operator of this equation acts in the Hilbert space $L^{2}_{(3)}({\mathbb R}^{3})$
of square-integrable vector-valued functions. This choice of the functional space is physically 
motivated by the energy considerations.
All components of the 
matrix-valued function $\chi(\bx,\omega)$ are presumed to be H{\"o}lder-continuous
in ${\mathbb R}^{3}$, i.e. not only inside the object, but also across its 
outer boundary. As previously $K$ denotes a compact operator,
which in this case is the last integral operator in (\ref{eq:EFVIE}) with 
weakly singular tensors (\ref{eq:Gten1})--(\ref{eq:AlpBetGam}) in its kernel.
Although, in general, weakly singular integral operators are not compact on ${\mathbb R}^{3}$,
the kernel of the last operator in (\ref{eq:EFVIE}) contains the contrast
function as well, which effectively restricts the spatial support of
the domain of the integral operator to $D$. Therefore, the total operator
is compact.

There exist other formulations of this problem. For example, for isotropic scatterers 
with differentiable constitutive parameters the regularization of the integral equation can 
be carried out explicitly, see \cite{Muller}, \cite{ColtonKress}. However, from the discussion  
of the previous sections we know that the spectra of the original and the regularized operator may be
very different.  

To find the spectrum of our operator we first need to derive its symbol.
The most laborous task here is, obviously, the Fourier transform of the simple singular integral operator.
The function $f(\bTheta)$ required in (\ref{eq:DeriveFk}) is in our case
 \begin{align}
 \label{eq:FThetaEM}
 f(\bTheta)=\frac{1}{4\pi}\left[3{\mathbb Q}-{\mathbb I}\right]\,.
 \end{align}
It is a matrix-valued function and therefore its Fourier image $F(\tilde{\bTheta})$ is
also a matrix-valued function. The complete symbol of the electromagnetic volume integral operator was 
first obtained in \cite{SamokhinBook} using the technique 
suggested in \cite{Mikhlin2}, where $f(\bTheta)$ is expanded in terms of 
spherical harmonics. Comparing the result of \cite{SamokhinBook} with the present formulation we
find that
 \begin{align}
 \label{eq:FkEM}
 F(\tilde{\bTheta})=\frac{1}{3}{\mathbb I} - \tilde{\mathbb Q}\,,
 \end{align}
where $\tilde{\mathbb Q}=\tilde{\bTheta}(\tilde{\bTheta}\cdot\;)$, or in matrix notation 
$\tilde{\mathbb Q}=\tilde{\bTheta}\tilde{\bTheta}^{T}$. Now using Definition~\ref{def:Symbol}
and expression (\ref{eq:ChiTen}) we obtain the symbol of the electromagnetic volume integral operator for an 
anisotropic scatterer as
 \begin{align}
 \label{eq:SymbolEM}
 \Phi(\bx,\tilde{\bTheta})={\mathbb I} + \tilde{\mathbb Q}\chi(\bx,\omega)={\mathbb I} - \tilde{\mathbb Q} + \tilde{\mathbb Q}\,\eta_{\rm r}(\bx,\omega)\,,
 \end{align}
where both $\chi$ and $\eta_{\rm r}=\eta/\eta_{\rm b}$ are matrix-valued functions and in general do not commute with matrix $\tilde{\mathbb Q}$.
The inverse of this symbol can be found explicitly as
 \begin{align}
 \label{eq:SymbolEMInv}
 \Phi^{-1}(\bx,\tilde{\bTheta})={\mathbb I} + \xi^{-1}\tilde{\mathbb Q} - \xi^{-1}\tilde{\mathbb Q}\,\eta_{\rm r}(\bx,\omega)\,,
 \end{align}
where
 \begin{align}
 \label{eq:Xi}
 \xi(\bx,\tilde{\bTheta})=\tilde{\bTheta}^{T}\eta_{\rm r}(\bx,\omega)\tilde{\bTheta}\,.
 \end{align}
In isotropic case, where $\eta(\bx,\omega)$ is a scalar, the inverse of the symbol
looks even simpler:
 \begin{align}
 \label{eq:SymbolEMInvIso}
 \Phi_{\rm isotropic}^{-1}(\bx,\tilde{\bTheta})={\mathbb I} - \tilde{\mathbb Q} + \tilde{\mathbb Q}\,\eta_{\rm r}^{-1}(\bx,\omega)
 ={\mathbb I}+\tilde{\mathbb Q}\chi'(\bx,\omega)\,.
 \end{align}
with $\chi'$ given by
 \begin{align}
 \label{eq:Chi'}
 \chi'(\bx,\omega)=\frac{\eta_{\rm b}(\omega)}{\eta(\bx,\omega)}-1\,.
 \end{align}
Although it is not our primary concern, now we can find 
the explicit electromagnetic regularizer. Notice that expression (\ref{eq:SymbolEMInvIso}) is identical in 
form to the symbol (\ref{eq:SymbolEM}) of the original operator. This means that the operator corresponding
to $\Phi^{-1}$ has the same form as the operator corresponding to $\Phi$. Hence in the isotropic case the regularizer 
is 
 \begin{align}
 \label{eq:RegEM}
 A^{\rm r}_{\rm isotropic}={\rm d}({\mathbb I}+\frac{1}{3}\chi') + A^{\rm s}{\rm d}(\chi')\,,
 \end{align}
with $\chi'$ given by (\ref{eq:Chi'}).

The essential spectrum consists of the eigenvalues of the symbol matrix $\Phi(\bx,\tilde{\bTheta})$ 
for all $\bx\in{\mathbb R}^{3}$ and all $\tilde{\bTheta}$ on the unit sphere ${\mathbb S}$. 
Direct computaions show that in the anisotropic case these 
eigenvalues are $\lambda_{1}=\lambda_{2}=1$, and 
$\lambda_{3}=\tilde{\bTheta}^{T}\eta_{\rm r}(\bx,\omega)\tilde{\bTheta}$,
and they depend on both $\bx$ and $\tilde{\bTheta}$. Whereas in the isotropic case, 
where $\eta_{\rm r}$ is a scalar, the eigenvalues are $\lambda_{1}=\lambda_{2}=1$
and $\lambda_{3}=\eta_{\rm r}(\bx,\omega)$, and they depend on $\bx$ only. 
Hence in general the essential spectrum is given by:
 \begin{align}
 \label{eq:EssSpecAnisotropic}
 \lambda^{\rm ess}_{1}=\lambda^{\rm ess}_{2}=1\,,\;\;\;\;\;\;\lambda^{\rm ess}_{3}=\tilde{\bTheta}^{T}\eta_{\rm r}(\bx,\omega)\tilde{\bTheta}\,,\;\;\;\;
 \bx\in{\mathbb R}^{3}\,,\;\;\;\;\tilde{\bTheta}\in{\mathbb S}\,.
 \end{align}
Notice that there is also a parameteric dependence on the angular frequency $\omega$
in both isotropic and anisotropic cases, which matters if the scatterer is conductive and/or dispersive.

The object parameters are presumed to be H{\"o}lder continuous functions of $\bx$ on ${\mathbb R}^{3}$.
This means that even with a homogeneous scatterer, characterized by $\eta_{\rm r}(\bx,\omega)={\rm const}$, $\bx\in D$, 
there exists a smooth transition between $\eta_{\rm r}$ and $1$ -- the background medium parameter.
Then, the essential spectrum must connect $1$ and point $\eta_{\rm r}$ in the complex plane.
Morover, for any inhomogeneous scatterer the essential spectrum will emerge from point $1$ on the real axis,
and continuously span all values admitted by $\tilde{\bTheta}^{T}\eta_{\rm r}(\bx,\omega)\tilde{\bTheta}$ 
for $\bx\in D$ and $\tilde{\bTheta}\in{\mathbb S}$.

Apart from the essential spectrum our operator may have discrete eigenvalues, which correspond
to nontrivial solutions of the problem
 \begin{align}
 \label{eq:EigenProblem}
 \left[{\rm d}({\mathbb I}+\frac{1}{3}\chi) + A^{\rm s}{\rm d}({\chi}) + K\right] u -\lambda u = 0\;.
 \end{align}
We shall consider here only the isotropic case.
In this respect it is convenient to use the standard sufficient uniqueness condition for the original problem
(\ref{eq:EFVIEOp}), which in terms of the admittance function $\eta(\bx,\omega)$ given by (\ref{eq:EtaTen}) can be written as
 \begin{align}
 \label{eq:Unique}
 {\rm Re}\left[\eta(\bx,\omega)\right] > 0, \;\;\; \bx \in D.
 \end{align}
Under this condition the null-space of the volume integral operator (operator in square brackets in (\ref{eq:EigenProblem}))
is known to be trivial. Considering the case where $\lambda\ne 1$ we can re-write  (\ref{eq:EigenProblem})
as 
 \begin{align}
 \label{eq:EigenProblemMod}
 \left[{\rm d}({\mathbb I}+\frac{1}{3}\chi') + A^{\rm s}{\rm d}({\chi'}) + K'\right] u = 0\;,
 \end{align}
where the only difference with the original operator is the new contrast function $\chi'$ given by
 \begin{align}
 \label{eq:NewChi}
 \chi'(\bx,\omega,\lambda)=\frac{\chi(\bx,\omega)}{1-\lambda}=
 \frac{\eta(\bx,\omega)- \lambda\eta_{\rm b}(\omega)}
 {\eta_{\rm b}(1-\lambda)}-1=
 \frac{\eta'(\bx,\omega,\lambda)}{\eta_{\rm b}(\omega)}-1\,.
 \end{align}
And since the eigenvalue problem has been reduced to the problem about the non-trivial null-space 
of the original operator we can use an `inverse' of the condition (\ref{eq:Unique}), i.e.
 \begin{align}
 \label{eq:BoundChiNew}
 {\rm Re}\left[\eta'(\bx,\omega,\lambda)\right] \le 0, \;\;\; \bx \in D.
 \end{align}
From here we derive the following bound on the eigenvalues:
 \begin{align}
 \label{eq:EigGen}
 \begin{split}
 {\rm Re}\,\eta(\bx,\omega) &-
 \left[{\rm Re}\,\eta(\bx,\omega) + {\rm Re}\,\eta_{\rm b}(\omega)\right] {\rm Re}\lambda
 \\
 &- \left[{\rm Im}\,\eta(\bx,\omega)-{\rm Im}\,\eta_{\rm b}(\omega)\right]{\rm Im}\lambda
 + {\rm Re}\,\eta_{\rm b}(\omega)\vert\lambda\vert^{2} \le 0,
 \;\;\bx\in D.
 \end{split}
 \end{align}
In addition we note that
 \begin{align}
 \label{eq:SpecRad}
 \vert\lambda\vert \le \Vert A \Vert < \infty,
 \end{align}
since $A$ is bounded, and that $\lambda=0$ is not an eigenvalue if the scattering 
problem has a unique solution (which we presume from now on).

The only natural {\it infinite} homogeneous background medium is vacuum, where $\sigma_{\rm b}(\omega)=0$
and $\varepsilon_{\rm b}(\omega)=\varepsilon_{0}$. 
In particular cases, such as scattering from subsurface objects, other homogeneous 
backgrounds can be used as an approximation. We shall deal with those later. 
Let us first consider the natural case, and in addition presume that $\sigma(\bx,\omega)=\sigma(\bx)$ 
and $\varepsilon(\bx,\omega)=\varepsilon(\bx)$, and are real-valued functions. 
Further we also assume that the contrast in permittivity 
$\varepsilon(\bx)-\varepsilon_{\rm b}$ is positive for all $\bx\in D$.
These assumptions hold for a static model of a conducting dielectric, which is applicable 
at low frequencies where the dispersion of constitutive paremeters is insignificant. 
In this case the eigenvalue bound (\ref{eq:EigGen}) reduces to
 \begin{align}
 \label{eq:EigGenReal}
 \begin{split}
 \sigma(\bx) - \sigma(\bx) {\rm Re}\lambda
 + \omega\left[\varepsilon(\bx)-\varepsilon_{0}\right]{\rm Im}\lambda \le 0,
 \;\;\bx\in D.
 \end{split}
 \end{align}
\begin{figure}[t]
\centerline{\epsfig{file=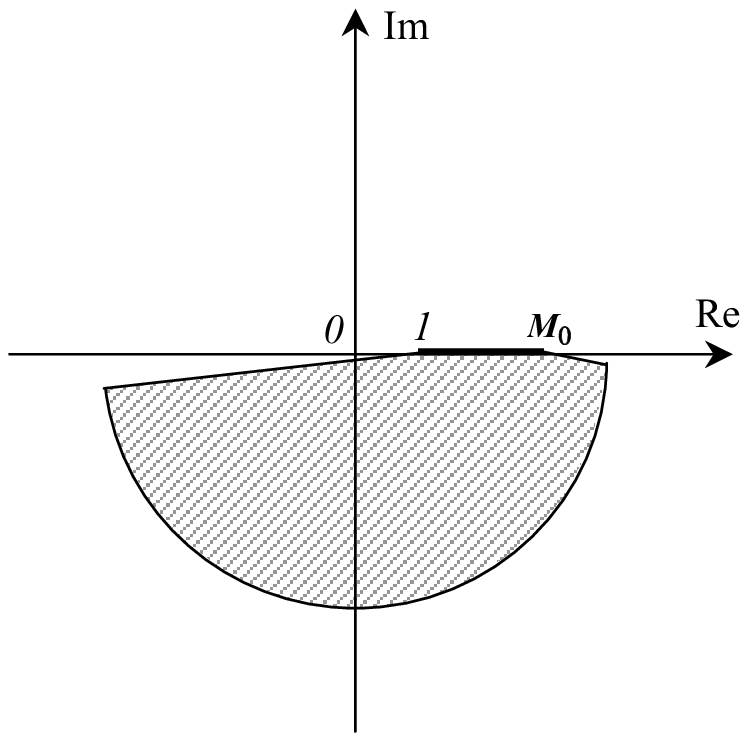,width=0.5\columnwidth,height=0.5\columnwidth}
            \epsfig{file=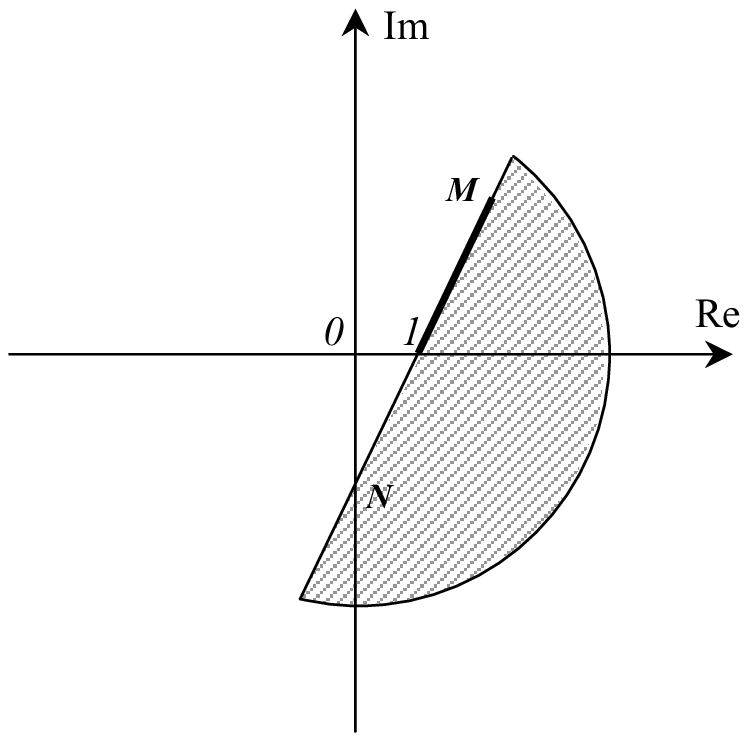,width=0.5\columnwidth,height=0.5\columnwidth}}
\caption{Location of the spectrum for a homogeneous object in vacuum. 
Left: nonconducting dielectric. 
Right: conducting dielectric.}
\end{figure}
In Fig.~5.1 and Fig.~5.2 (left) we present geometric estimates derived from this formula 
for the following three situations:
 \begin{align}
 \label{eq:Inh1}
 \sigma(\bx)& = 0,\;\;\;\;\;\;\;\;\;\;\;\;\;\;\; 1<\frac{\varepsilon(\bx)}{\varepsilon_{\rm 0}}\le M_{0}; 
 \\
 \label{eq:Hom1}
 \sigma(\bx)& = {\rm const} > 0,\;\; \varepsilon(\bx)={\rm const}>\varepsilon_{\rm 0}; 
 \\
 \label{eq:Inh2}
 \sigma(\bx)& > 0,\;\;\;\;\;\;\;\;\;\;\;\;\;\;\; \varepsilon(\bx)>\varepsilon_{\rm 0}; 
 \end{align}
where $\bx\in D$. 
The other parameters in these figures are given by
 \begin{align}
 \label{eq:M}
 M&= \frac{\varepsilon}{\varepsilon_{0}} + i \frac{\sigma}{\omega\varepsilon_{0}},
 \;\;\;
 N=- i \frac{\sigma}{\omega(\varepsilon-\varepsilon_{0})},
 \\
 \label{eq:N1}
 N_{1}&=- i \max\limits_{\bx\in D}\frac{\sigma(\bx)}{\omega\left[\varepsilon(\bx)-\varepsilon_{0}\right]},
 \\
 \label{eq:N2}
 N_{2}&=- i \min\limits_{\bx\in D}\frac{\sigma(\bx)}{\omega\left[\varepsilon(\bx)-\varepsilon_{0}\right]},
 \end{align}
Fat solid lines and curves schematically outline the essential part of the spectrum,
which in the case of a homogeneous object we presume to be straight lines
connecting the real $1$ and the corresponding value of $\eta/\eta_{\rm b}$.
As one can see, in the natural (vacuum) background medium, the bounds are wedge shaped, same as in \cite{Kleinman1990}.
In the case of an object with losses these bounds safely separate 
the domain of eigenvalues from the zero of the complex plane. This is important since 
the uniqueness condition can only guarantee that there are no eigenvalues {\it equal} to zero, 
but does not tell if there are any eigenvalues {\it close} to zero, while such eigenvalues 
may cause instabilities in the numerical solution and slow down the convegrence of an iterative 
algorithm. From this point of view the worst case 
is the lossless object or an inhomogeneous object with lossless parts. 

Now let us analyze the popular approximation, where the homogeneous 
background medium is considered to be conducting, i.e. $\sigma_{\rm b} > 0$ and/or has arbitrary 
constant permittivity $\varepsilon_{\rm b}>\varepsilon_{0}$. Such approximations are often
used if a scatterer is located inside a large but finite and more or less homogeneous 
host medium, say, inside the Earth or a human body, and the reflections from the outer 
boundary of the host medium can be neglected.
We again consider the static model of a dielectric and from (\ref{eq:EigGen}) we 
derive the following estimate
 \begin{align}
 \label{eq:LossBack}
 \left[{\rm Re}\lambda - \frac{A}{2}\right]^{2} + \left[{\rm Im}\lambda + \frac{B}{2}\right]^{2} < \left(\frac{D}{2}\right)^{2},
 \end{align}
which shows that all eigenvalues are situated inside a circle.
Parameters of this circle's center are 
 \begin{align}
 \label{eq:AB}
 A=\frac{\sigma(\bx)}{\sigma_{\rm b}}+1,
 \;\;\;
 B=\frac{\omega}{\sigma_{\rm b}}\left[\varepsilon(\bx)-\varepsilon_{\rm b}\right],
 \end{align}
and its diameter can be found from
 \begin{align}
 \label{eq:D}
 D^{2}=\left[\frac{\sigma(\bx)}{\sigma_{\rm b}}-1\right]^{2} + 
 \omega^{2}\left[\frac{\varepsilon(\bx)-\varepsilon_{\rm b}}{\sigma_{\rm b}}\right]^{2}.
 \end{align}
\begin{figure}[t]
\centerline{\epsfig{file=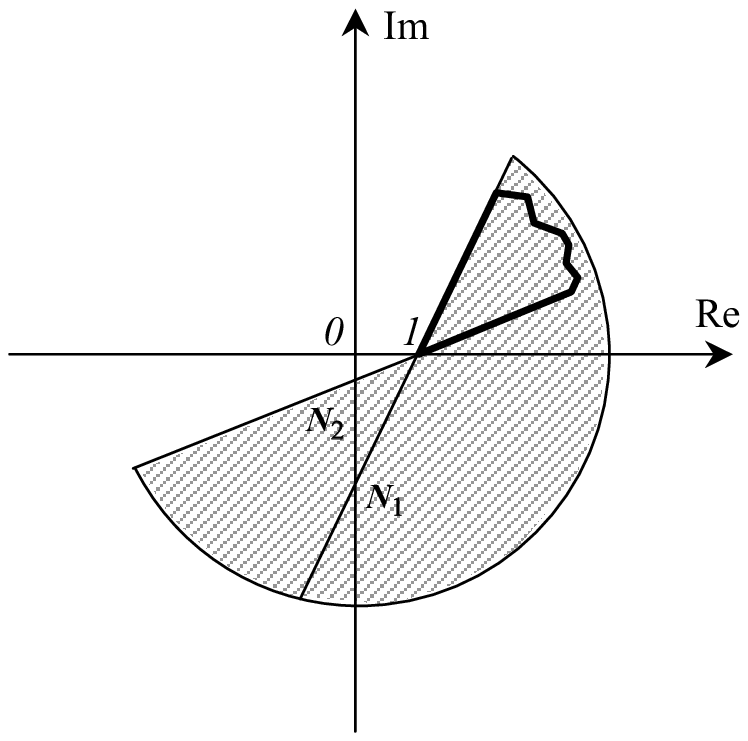,width=0.5\columnwidth,height=0.5\columnwidth}
            \epsfig{file=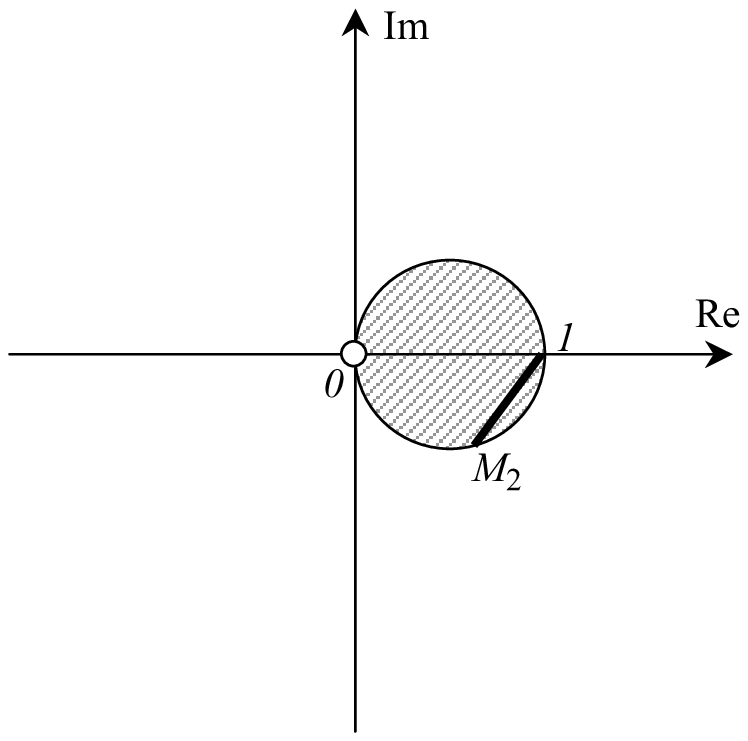,width=0.5\columnwidth,height=0.5\columnwidth}}
\caption{Location of the spectrum. Left: Inhomogeneous conducting dielectric in vacuum.
Right: Homogeneous nonconducting dielectric object in a conducting dielectric background 
(no contrast in permittivity). }
\end{figure}
Figure~5.2 (right) corresponds to a particular situation where a nonconducting 
object ($\sigma(\bx)=0$) with no contrast in permittivity ($\varepsilon(\bx)=\varepsilon_{\rm b}$) 
is immersed in a conducting background. This can represent an air pocket or a low-contrast 
plastic landmine in wet soil. The constant $M_{2}$ is given by
 \begin{align}
 \label{eq:M2}
 M_{2}=\frac{\eta}{\eta_{\rm b}}=\frac{\omega^{2}\varepsilon_{\rm b}^{2}}{\sigma_{\rm b}^{2} + \omega^{2}\varepsilon_{\rm b}^{2}}
 - i \frac{\omega\varepsilon_{\rm b}\sigma_{\rm b}}{\sigma_{\rm b}^{2} + \omega^{2}\varepsilon_{\rm b}^{2}}.
 \end{align}
On one hand, the spectrum is now explicitly bounded allowing, in particular,
for a more precise convergence estimate when an iterative method is employed 
to solve the scattering problem. On the other hand, the circle extends towards 
the zero of the complex plane as in Fig.~5.2 (right), 
meaning that, although there is no zero eigenvalue in the spectrum, 
some of the eigenvalues may get close to zero, and render
the scattering problem unstable. 

The last point to note is about the relative weight of the compact and singular parts
in the operator of (\ref{eq:EFVIE}) as a function of angular frequency $\omega$.
The coefficients $\alpha$ and $\beta$ in (\ref{eq:AlpBetGam}) depend
on the wavenumber $k_{\rm b}(\omega)$, which in its turn is proportional 
to $\omega$. Hence, the kernel of the compact operator gets more `weight'
as frequency increases, and the norm of the operator increases as well. 
As the compact operator delivers only eigenvalues, we can expect that the 
eigenvalues will spread out at higher frequencies.

\section{Numerical experiments}
There is only a limited correspondence between the spectra of an integral operator
and of its discretized (matrix) version. First of all, matrices do not
have continuous spectra, but only discrete eigenvalues. Hence, we should not
expect to see the lines and curves of the previous section. On the other hand,
the continuous spectrum may serve as an accumulation area for the eigenvalues,
as the latter should converge to the spectrum of an operator in the continuous limit.
Let us first see if it is indeed so. 

We discretize the operator of (\ref{eq:EFVIE}) using the standard collocation 
method (see e.g. \cite{Rahola2000}, \cite{SamokhinBook}), which gives an order $h^{2}$ accuracy, where $h$ is the size of
an elementary cubic cell. The scatterer is discretized on an $N$-point homogeneous grid,
and the resulting matrix is $3N\times 3N$ and is completely filled with complex 
numbers. The matrix eigenvalues are computed using the standard Matlab function ${\tt eig}$.

According to our theoretical predictions the essential spectrum, continuous or not,
dominates at lower frequencies (small objects). Therefore, we shall first consider
an object whose extent is sufficiently small. 
\begin{figure}[t]
\centerline{\epsfig{file=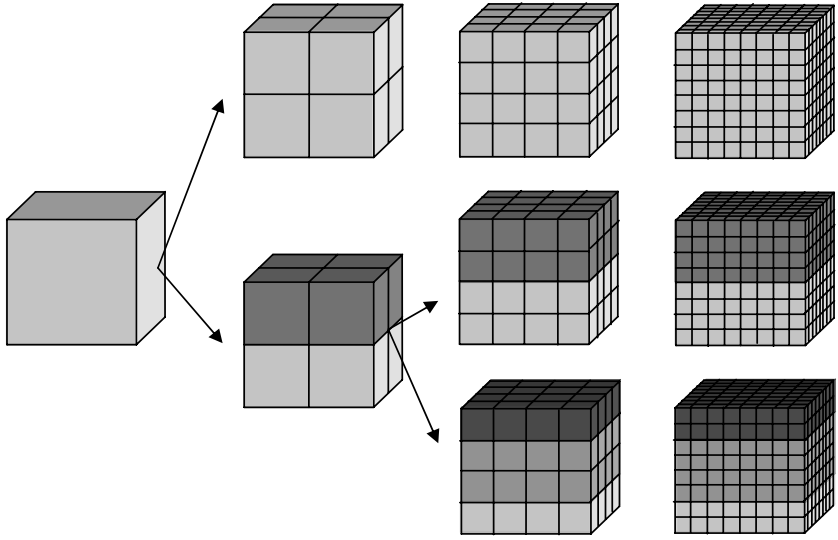,width=0.9\columnwidth,height=0.6\columnwidth}}
\caption{Different discretization (and quantization) levels for three cubes (homogeneous, two-, and three-layered). 
The side length of all cubes is $a\approx\lambda_{\rm med}/20$. Inhomogeneity becomes `visible'
only at an appropriate discretization level.
}
\end{figure}
\begin{figure}[t]
\centerline{\epsfig{file=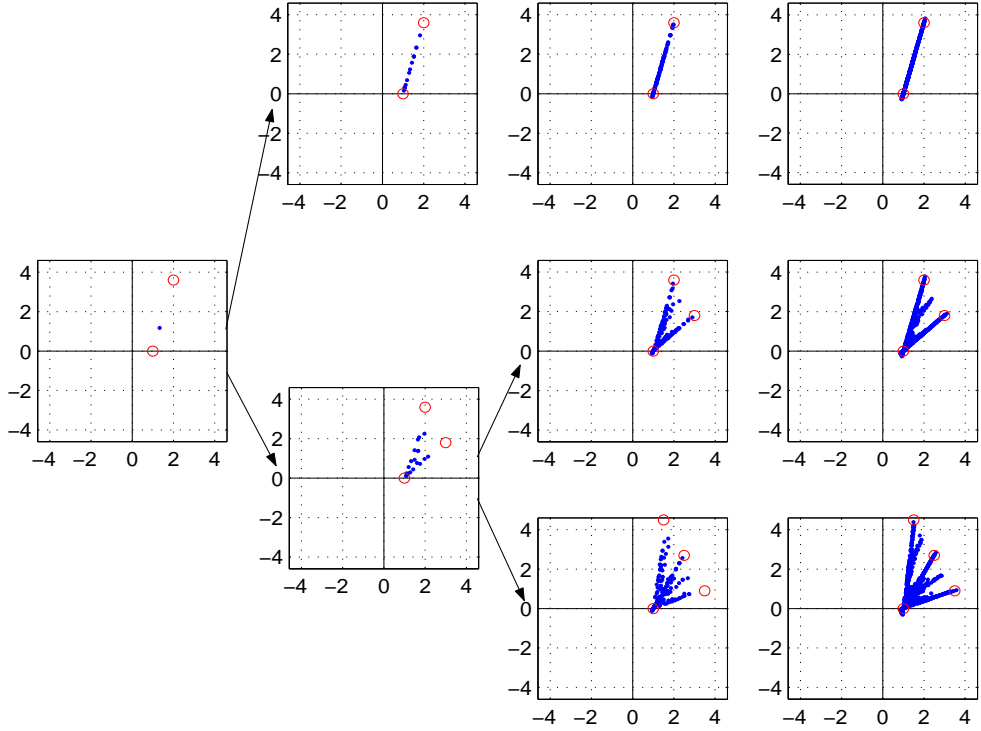,width=\columnwidth,height=0.75\columnwidth}}
\caption{Effect of discretization (and quantization) refinement on the matrix eigenvalues in the 
low-frequency (small object) regime. Plots correspond to the objects shown in Fig.~6.1, 
constitutive parameters $\eta/\eta_{\rm b}$ are given as circles. 
}
\end{figure}
The background medium is vacuum and the frequency is set to $\omega/(2\pi)=1$~GHz.
The scatterer has the shape of a cube with side length $a\approx(\lambda_{\rm med}/20)$,
where $\lambda_{\rm med}$ is the medium wavelength, see Fig.~6.1. According to the common
{\it ad hoc} discretization rule such a small object can be 
modelled as a single cell (Fig.~6.1, left). This, of course, would only give an order $h^{2}$ approximation
of the field at the geometrical center of the scatterer. In an attempt to compute the field
at other points inside the object one may wish to introduce a finer discretization.
Then, if the object is homogeneous, i.e. $\eta(\bx)={\rm const}$, the value of 
the consitutive parameter would be the same at all internal grid points (Fig.~6.1, upper row). 
If, however, the object is inhomogeneous, then the matrix will include  
the new, refined, grid values of $\eta(\bx_{n})$, $n=1,\dots,N$ (Fig.~6.1, middle and lower rows). 
Obviously, there is a link 
between the spatial discretization and the quantization of constitutive parameters, and
we have observed an interesting phenomenon related to these two processes 
in our numerical experiments. Namely, if we refine the spatial discretization while 
keeping constitutive parameters constant, then the (low-frequency) spectrum
converges to a set of very dense line segments connecting the real unit and
the corresponding values of $\eta/\eta_{\rm b}$. In this way we seem to model
an object consisting of one or more homogeneous parts without really telling 
what is the behavior of the constitutive parameter across the interfaces.
If, on the other hand, the quantization of the constitutive
parameter results in new values at a finer level of discretization, then 
the line segments in the spectrum are less pronounced.

Consider, for example the case of three different cubes having the same 
size, and the following sequence of grid steps: $h=a, a/2, a/4, a/8$. The matrix
dimensions are then, correspondingly: $3\times 3$, $24\times 24$, $192\times 192$, and $384\times 384$.
The objects and the eigenvalues
are presented, respectively, in Fig.~6.1 and Fig.~6.2. 
The upper row in both figures corresponds to the model of a completely homogeneous cube,
the middle row models a cube consisting of two layers, and the bottom row corresponds to a 
three-layered cube. The layered structure becomes visible at $h=a/2$ (second column), and the difference between
the two and the three layers is only seen at $h=a/4$ (third column). At $h=a/8$ (last column) we had no more 
changes in the constitutive parameters of all three objects. 
The constitutive parameters $\eta/\eta_{\rm b}$ are shown as circles in Fig.~6.2.
The dense lines appear as soon as we stop
refining the quantization, but keep refining the spatial discretization.
\begin{figure}[t]
\centerline{\epsfig{file=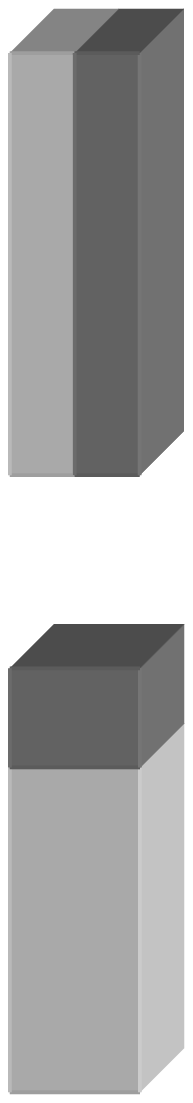,width=0.15\columnwidth,height=0.65\columnwidth}
\epsfig{file=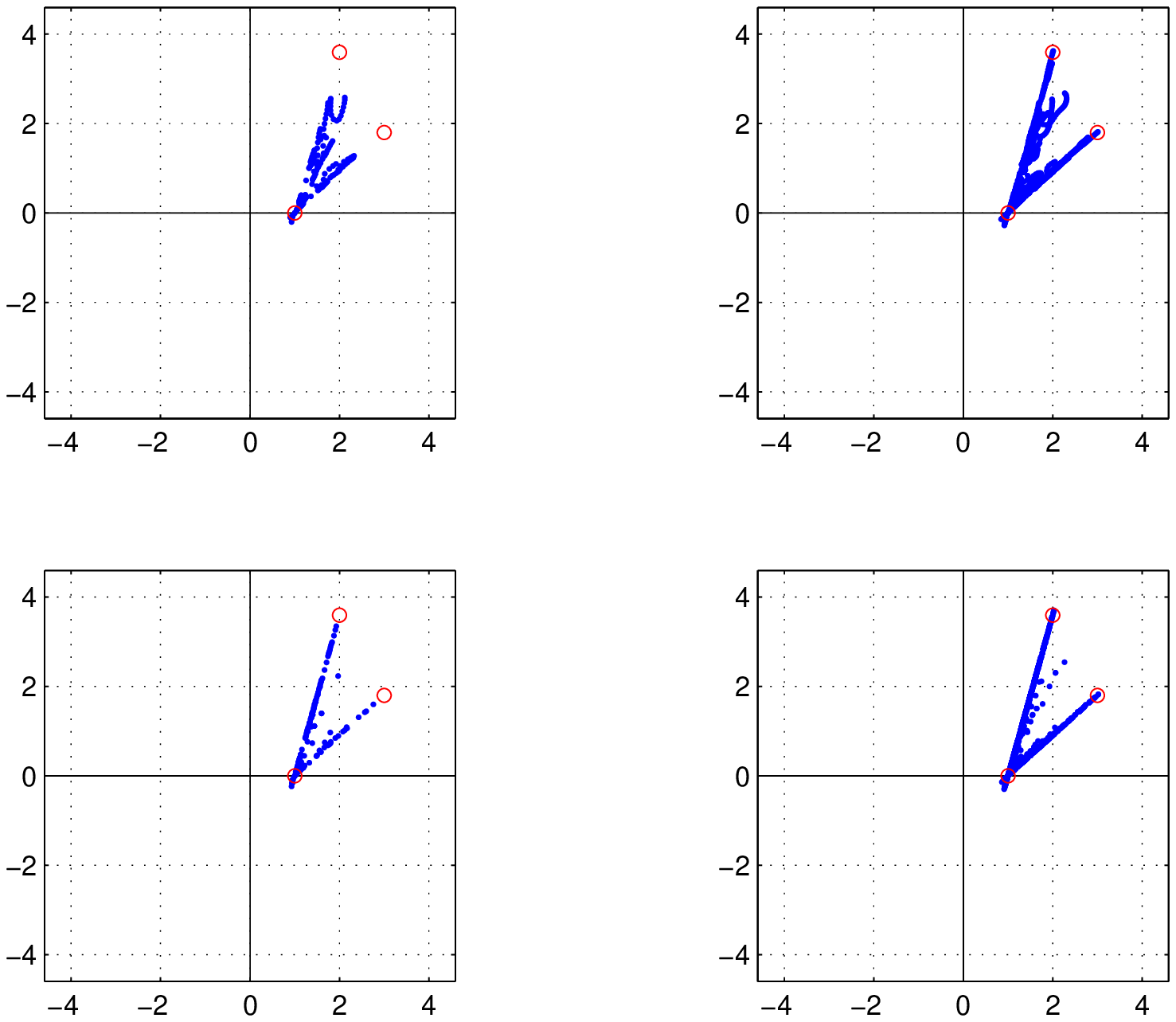,width=0.8\columnwidth,height=0.65\columnwidth}}
\caption{Effect of the object's shape on the matrix eigenvalues in the low-frequency (small object) regime:
same constitutive parameters $\eta/\eta_{\rm b}$ (circles), but different shapes.
From left to right: pictures of the objects, and eigenvalues for two discretization levels: $h=\lambda_{\rm med}/70$ and
$\lambda_{\rm med}/140$.
}
\end{figure}
Figure~6.3 illustrates the marginal influence of the object's geometry on the distribution of eigenvalues at
low frequencies (same frequency as before). Both objects (top and bottom row) consist of two 
homogenous parts with constitutive parameters given by circles and have the same outer shape 
(long parallelepiped, roughly $\lambda/5\times \lambda/42\times \lambda/42$). The difference between the 
objects is in the geometry of the parts. The upper row in Fig.~6.3 corresponds to the object consisting of two equal 
parts (roughly $\lambda/5\times \lambda/42\times \lambda/85$ each), which divide the parallelepiped {\it along} its
longer dimension.
Whereas, the lower row corresponds to the object consisting of two unequal 
parts (roughly $\lambda/21\times \lambda/42\times \lambda/42$ and $\lambda/7\times \lambda/42\times \lambda/42$)
dividing the parallelepiped {\it across} its longer dimension.
From left to right two discretization levels are presented: $h=\lambda_{\rm med}/70$ and $h=\lambda_{\rm med}/140$.
With both objects we notice the appearence of the line segments connecting the real unit and the values 
of  $\eta/\eta_{\rm b}$. The difference seems to be among the other eigenvalues situated between the line segments.
These and many other low-frequency numerical experiments suggest that the line segments observed here and 
in \cite{Rahola2000} are the matrix analogue of the operator's essential spectrum, and that they 
depend only on the grid values of $\eta(\bx)/\eta_{\rm b}$.

\begin{figure}[t]
\centerline{\epsfig{file=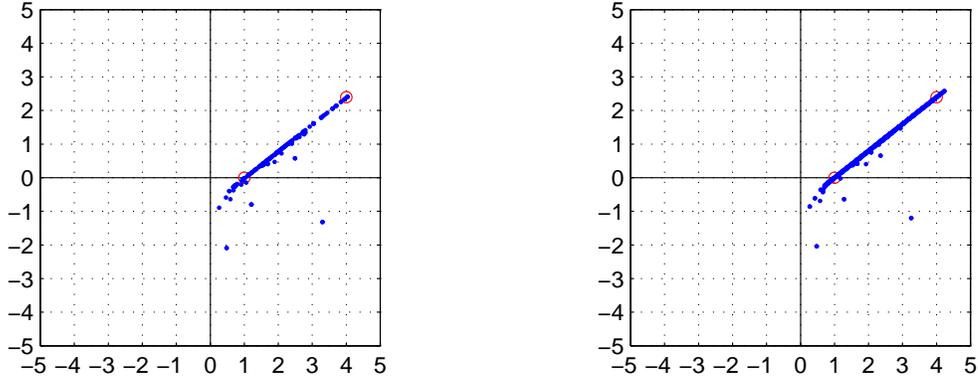,width=\columnwidth,height=0.4\columnwidth}}
\caption{Effect of discretization on the matrix eigenvalues at higher frequencies. 
Object - homogeneous cube.
Left: discretization level $h=\lambda_{\rm med}/5$. Right: discretization level $h=\lambda_{\rm med}/10$.
}
\end{figure}
Figure~6.4 gives the spectrum for a homogeneous cube at higher frequencies. The $\eta/\eta_{\rm b}$ 
of the object is shown as a circle. The length of the cube's side is chosen to coinside with the medium
wavelength in this case, i.e. $a=\lambda_{\rm med}$. One can see that now we have not only the dense line segment as in Fig.~6.2 (top-right),
but also a few off-line eigenvalues, which appear within the predicted bounds of Fig.~5.1 (right).
Two discretizations are shown: $h=\lambda_{\rm med}/5$ (Fig.~6.4 left),
and $h=\lambda_{\rm med}/10$ (Fig.~6.4 right). Normally, neither of these would be considered a ``proper'' level 
of discretization. However, here we observe an interesting phenomenon related to the essential spectrum,
which partly justifies a much more relaxed attitude of practitioners to the discretization of the electromagnetic volume integral equation, 
as opposed to the tough requirements on the discretization of differential Maxwell's equations via the finite-difference approach.
With $h=\lambda_{\rm med}/5$ we have in total 375 eigenvalues most of which are 
clustered along the line segment of the essential spectrum. With $h=\lambda_{\rm med}/10$ we already 
have 3000 eigenvalues, however, all `new' 2625 eigenvalues keep filling the same line segment. Of course, the off-line
eigenvalues do shift a little, but their number seems to remain the same for both discretization levels. This and similar experiments
at higher (resonance) frequencies show that the refinement of discretization has no significant effect on
the spectral radius of the matrix, which is what is expected from an integral equation formulation.

We conclude our numerical experiments with the result corresponding to the conductive background case
described in Fig.~5.2 (right). As one can see from Fig.~6.5 (left) the matrix spectrum is, indeed, quite accurately 
described by the predicted circle.
\begin{figure}[t]
\centerline{\epsfig{file=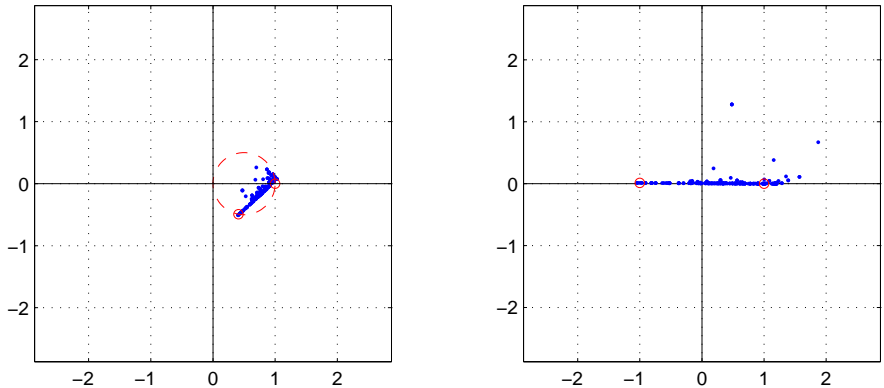,width=0.9\columnwidth,height=0.4\columnwidth}}
\caption{Spectrum in exotic cases. Left: vacuum cube in a lossy background medium. 
Right: cube with negative permittivity and small losses in vacuum background.
}
\end{figure}

\section{Conclusions and possible applications}
The spectrum of the volume integral operator of three-dimensional electromagnetic scattering contains 
both the essential continuous part and discrete eigenvalues. The apparent difference with the one- and two-dimensional 
electromagnetics, where the spectrum is purely discrete, stems from the presence of a strong singularity in the 
kernel of the integral operator in the three-dimensional case.
We have shown that the essential spectrum is explicitly given by 
(\ref{eq:EssSpecAnisotropic}) for any anizotropic scatterer with H{\"o}lder continuous constitutive parameters. 
Knowing the spectrum one can find, for instance, the relaxation parameter which provides the optimal convergence of the over-relaxation
iterative algorithm. Due to the fact that the well-described essential spectrum dominates at lower frequencies we suggest that
the over-relaxation method (which is very cheap from the computational viewpoint) should be used with 
quasi-static problems.

However, there exists a much more exciting direct application of this knowledge as well.
For some reason various electromagnetically `exotic' media are at the core of the present day research.
Take for example left-handed materials also known as media with negative refractive index.
These materials are supposed to be highly dispersive, so that for a certain range of $\omega$'s 
both the dielectric permittivity and the magentic permeability happen to have negative real parts.  
Although we have not discussed the magnetic case here, our preliminary calculations indicate that all the present 
conclusions hold, and that the magnetic properties of the object result in another similar contribution to the essential 
spectrum. Then, the spectrum of the `left-handed' object will not only contain points in the left part of the complex plane, 
but also lines or curves connecting these points with the real unit. For a low loss material, which for obvious reasons is the ultimate 
goal of experimentalists, the lines of essential spectrum may then proceed dengerously close to the zero of the complex plane as
shown in Fig.~6.5 (right), where $\epsilon/\epsilon_{\rm b}=-1$ and $\sigma=0.001$~S/m. 
From the physical viewpoint this would mean that the electromagnetic field is unstable in such media. 
Without losses ($\sigma=0$) 
the line would go right through the zero, meaning that the solution (electromagnetic field) does not exist at all.
May be that is the reason why we do not observe many left-handed substances in Nature?
Another interesting application where the explicit knowledge of the essential spectrum may be of help is
magnetically confined fusion plasma.

\section*{Acknowledgements}
This research is supported by the grant of the Netherlands Organization for Scientific Research (NWO)
and the Russian Foundation for Basic Research (RFBR).

\end{document}